\documentclass{article}
\usepackage{graphicx} 
\usepackage{caption,subcaption}
\usepackage{authblk}
\usepackage{hyperref}

\edef\restoreparindent{\parindent=\the\parindent\relax}
\usepackage{parskip}
\restoreparindent

\usepackage[margin=1in]{geometry}

\usepackage[citestyle=numeric-comp,bibstyle=numeric, natbib=true, backend=biber,hyperref=true,maxbibnames=1,maxcitenames=1,sorting=none]{biblatex}
\DeclareUnicodeCharacter{2500}{\textendash}

\title{Performance Classification and Remaining Useful Life Prediction of Lithium Batteries Using Machine Learning and Early Cycle Electrochemical Impedance Spectroscopy Measurements}
\author[1]{Christian Parsons}
\author[1]{Adil Amin}
\author[1]{Prasenjit Guptasarma}
\affil[1]{Department of Physics, University of Wisconsin–Milwaukee, Wisconsin, United States of America}
\date{August 2024}

\begin{document}

\maketitle

\section{Abstract}
We presents an approach for early cycle classification of lithium-ion batteries into high and low-performing categories, coupled with the prediction of their remaining useful life (RUL) using a linear lasso technique. Traditional methods often rely on extensive cycling and the measurement of a large number of electrochemical impedance spectroscopy (EIS) frequencies to assess battery performance, which can be time and resource consuming. In this study, we propose a methodology that leverages specific EIS frequencies to achieve accurate classification and RUL prediction within the first few cycles of battery operation. Notably, given only the 20 kHz impedance response, our support vector machine (SVM) model classifies batteries with 100\% accuracy. Additionally, our findings reveal that battery performance classification is frequency agnostic within the high frequency ($<20$ kHz) to low-frequency (32 mHz) range. Our model also demonstrates accurate RUL predictions with $R^2>0.96$ based on the out of phase impedance response at a single high (20 kHz) and a single mid-frequency (8.8 Hz), in conjunction with temperature data. This research underscores the significance of the mid-frequency impedance response as merely one among several crucial features in determining battery performance, thereby broadening the understanding of factors influencing battery behavior.

\section{Introduction}

The growing demand for electric vehicles \cite{maisel2023forecast,khan2023design}, residential energy storage systems \cite{fallahifar2023optimal,terlouw2023designing}, smart phones, and other portable devices \cite{adnan2023future} has led to a continual increase in the utilization of lithium-ion batteries.  Assessing the state of health (SOH) and predicting the remaining useful life (RUL) of lithium-ion batteries are pivotal in ensuring their efficient and reliable operation \cite{xiong2018towards,pradhan2022battery,tian2020review}. Traditional methods for evaluating battery performance often involve extensive cycling and the measurement of numerous electrochemical impedance spectroscopy (EIS) frequencies or charge/discharge voltage or capacity curves. This can be laborious, time-consuming, and resource-intensive \cite{wang2017battery, weng2013board,BERECIBAR2016239,ng2014naive,thelen2022augmented}. Accurate RUL classification and predictions during production enable manufacturers to quickly identify underperforming batteries, guaranteeing only the highest quality and safest products reach the market \cite{qian2021role}. Similarly, integrating straightforward and precise RUL predictions can empower academic labs to prioritize promising battery material candidates for further development and study. In this paper, we present an approach to perfectly accurate classification of lithium-ion batteries into high or low performing classes using a support vector machine model, and to accurately ($R^2>0.96$) predict the RUL of lithium-ion batteries using a lasso regression algorithm. 

Capacity degradation with cycling, which determines RUL, is caused by several key physical processes that are difficult to directly measure in real time without destroying the battery \cite{garg2019crystal}. Recent work has suggested that interfacial properties are the dominant source of capacity degradation in these metal-oxide lithium-ion batteries \cite{Zhang2020degradation}. However, other work  suggests a combination of several processes such as the formation of a solid electrolyte interface, cathode material expansion, contraction, cracking, dendrite formation, electrolyte degradation, and active material loss contribute to the overall degradation of the battery \cite{arora1998capacity, han2019review, lin2013comprehensive, zhang2000studies}. 

EIS  provides information about the internal electrochemical processes occurring within a battery cell in real time, without damaging or destroying the cell, via the impedance response to a small perterbation \cite{chang2010electrochemical}. The impedance response of the battery due to a small amplitude alternating current (AC) signal is measured over a range of frequencies, typically from mHz to kHz \cite{lazanas2023electrochemical}. This electrochemical impedance response can be interpreted and quantified via equivalent circuit models, where physical processes can be associated with elements of the equivalent circuit model \cite{Qu2022}. In a typical system the following associations can be made\cite{choi2020modeling, Qu2022}: the real impedance offset is due to electrolyte and seperator Ohmic resistance, the high frequency response is due to the formation of a solid electrolyte interphase (SEI) layer, the mid frequency response is due to charge transfer, and the low frequency response is due to diffusion of lithium ions at the electrodes. The large range of frequencies typical measured with EIS and used as inputs into many machine learning models requires specialized equipment to precisely measure\cite{esser2022electrochemical, lohmann2015electrochemical}. Additionally, significant computational resources may be required to train models on the large and high-dimensional datasets \cite{Zhang2020degradation, nguyen2019exact, sarker2021machine}. 

 In this paper, we present an approach that employs a support vector machine technique to classify lithium-ion batteries from the Zhang dataset \cite{Zhang2020degradation} into high and low performing categories within the initial cycles of operation. In contrast to conventional methods that rely on a large number of EIS frequencies, our proposed methodology focuses on identifying specific EIS frequencies that yield accurate classification results. Here we find that the 20 kHz impedance response gives $100 \%$ accurate classifications of high-performing batteries and low-performing batteries. High and low-performing batteries are respectively defined as having a cycle 200 capacity higher or lower than 80 \% of its initial capacity. We find that the model can be further extended to lower single frequencies, 8.8 Hz or 32 mHz, with a small reduction in classification accuracy. We also show that a linear lasso algorithm gives accurate predictions ($R^2 > 0.96$) of the RUL of these batteries with impedance response at 20 kHz and 8 Hz as well as temperature as our model features. We further show that our lasso model can be further simplified to only require the impedance response at 20 kHz and temperature with an ($R^2 > 0.90$).

Our study highlights the frequency-agnostic nature of battery performance classification in the high frequency (20 kHz) to mid-frequency (8 Hz) range, indicating that the effectiveness of our approach transcends specific EIS measurement frequencies. Additionally, we showcase the capability of accurate RUL predictions at early cycles by using only temperature and the out-of-phase impedance response at both a high and a mid-frequency. By underscoring the importance of mid-frequency impedance response as just one among several predictive features influencing battery performance, our work advocates for a holistic approach to improving battery performance. Furthermore, our demonstration of a broad range of predictive frequency responses offers battery researchers and manufacturers the flexibility to use convenient frequencies in their analyses.

\section{Results}
\subsection{Classification}

We begin by exploring the feasibility of accurately categorizing lithium-ion batteries into high-performing and low-performing classes within the initial few cycles with non-invasive EIS measurements. For the purposes of this paper, we define a high-performing battery, given label '1', as a battery which meets the condition $C_{200} > 0.8*C_{1}$. A low-performing battery meets the condition $C_{200} < 0.8*C_{1}$. Here, $C_{200}$ is the discharge capacity just prior to cycle 200 and $C_{1}$ is the discharge capacity measured at cycle 1. All analysis is done using the open-source Zhang lithium-ion EIS-Capacity dataset\cite{Zhang2020degradation}. To avoid biases, the definition of battery failure ($0.8C_1$) and our train-test split is equivalent to those used by Zhang. Batteries labeled 25C01-25C04, 35C01, and 45C01 make up the the training det while batteries labeled 25C05-25C08, 35C02, and 45C02 make up the the test set. Battery 25C04 in the training set is classified as low-performing based on linear extrapolation. Due to the limited number of independent batteries in this dataset, we favour a support vector machine (SVM) \cite{smola2004tutorial} algorithm for classification, described in Methods. 

\begin{figure}[htb]
    \centering 
\begin{subfigure}{0.33\textwidth}
  \includegraphics[width=\linewidth]{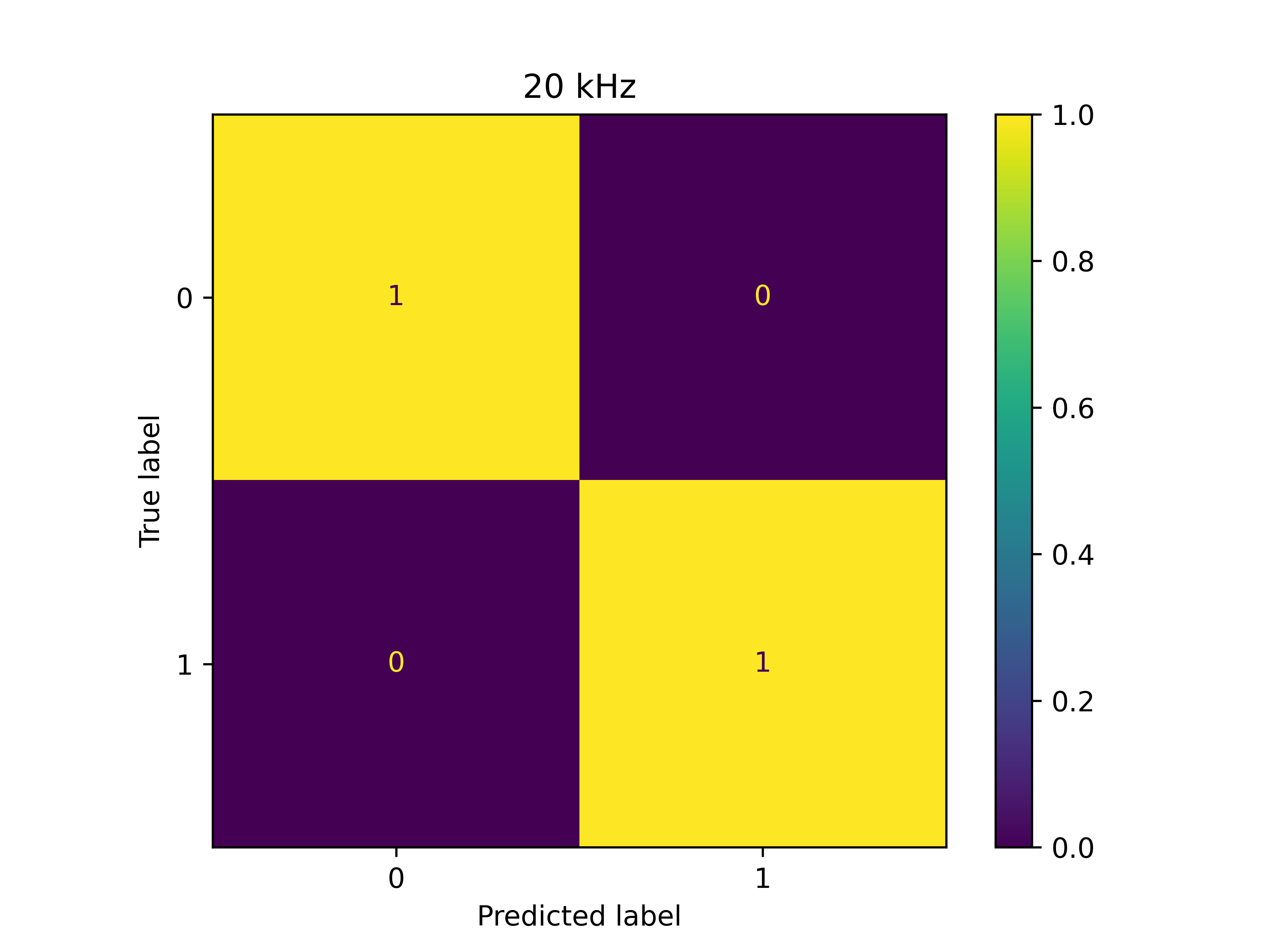}
  \caption{}
  \label{CMVa}
\end{subfigure}\hfil
\begin{subfigure}{0.33\textwidth}
  \includegraphics[width=\linewidth]{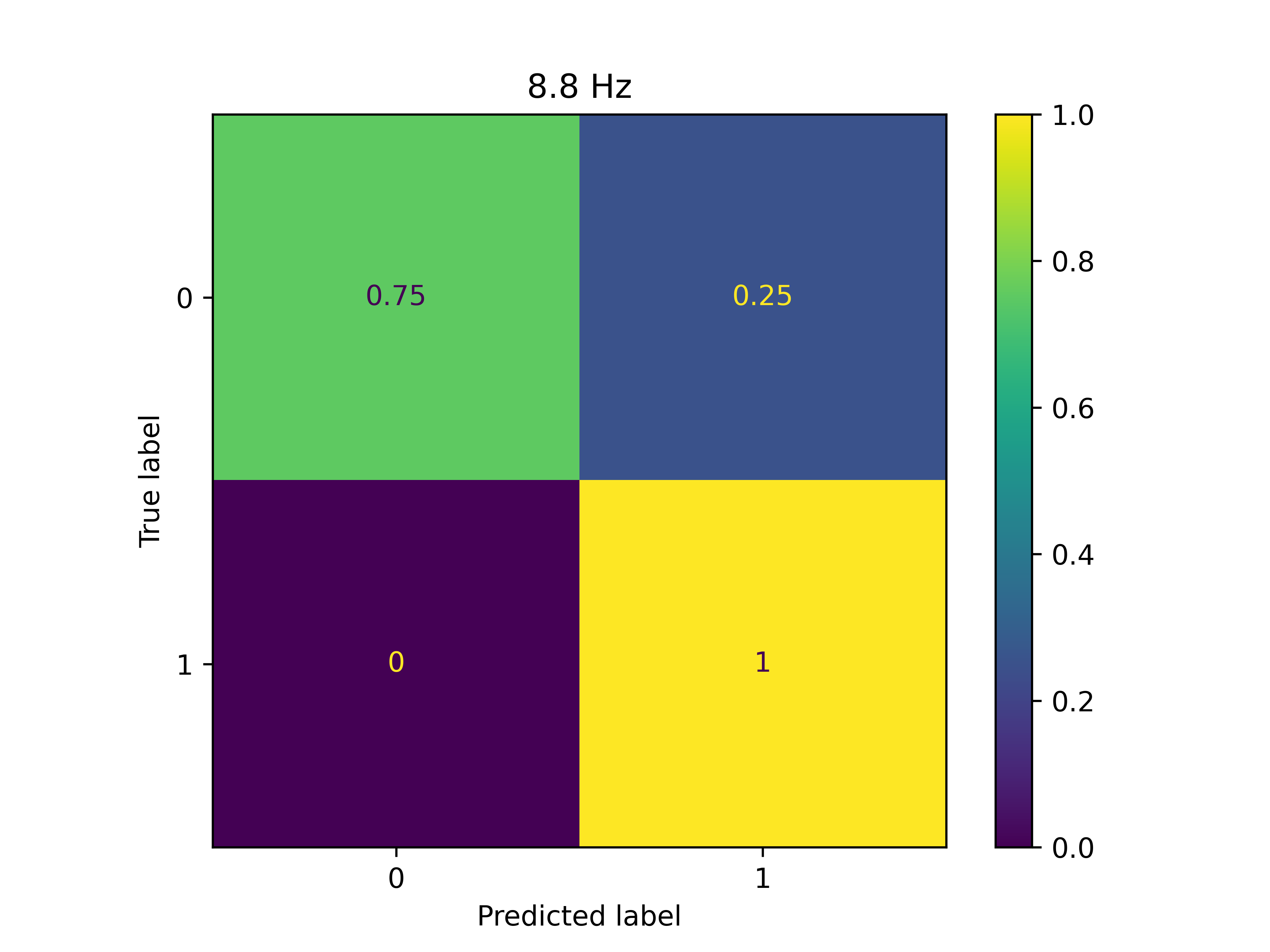}
  \caption{}
  \label{CMVb}
\end{subfigure}\hfil
\begin{subfigure}{0.33\textwidth}
  \includegraphics[width=\linewidth]{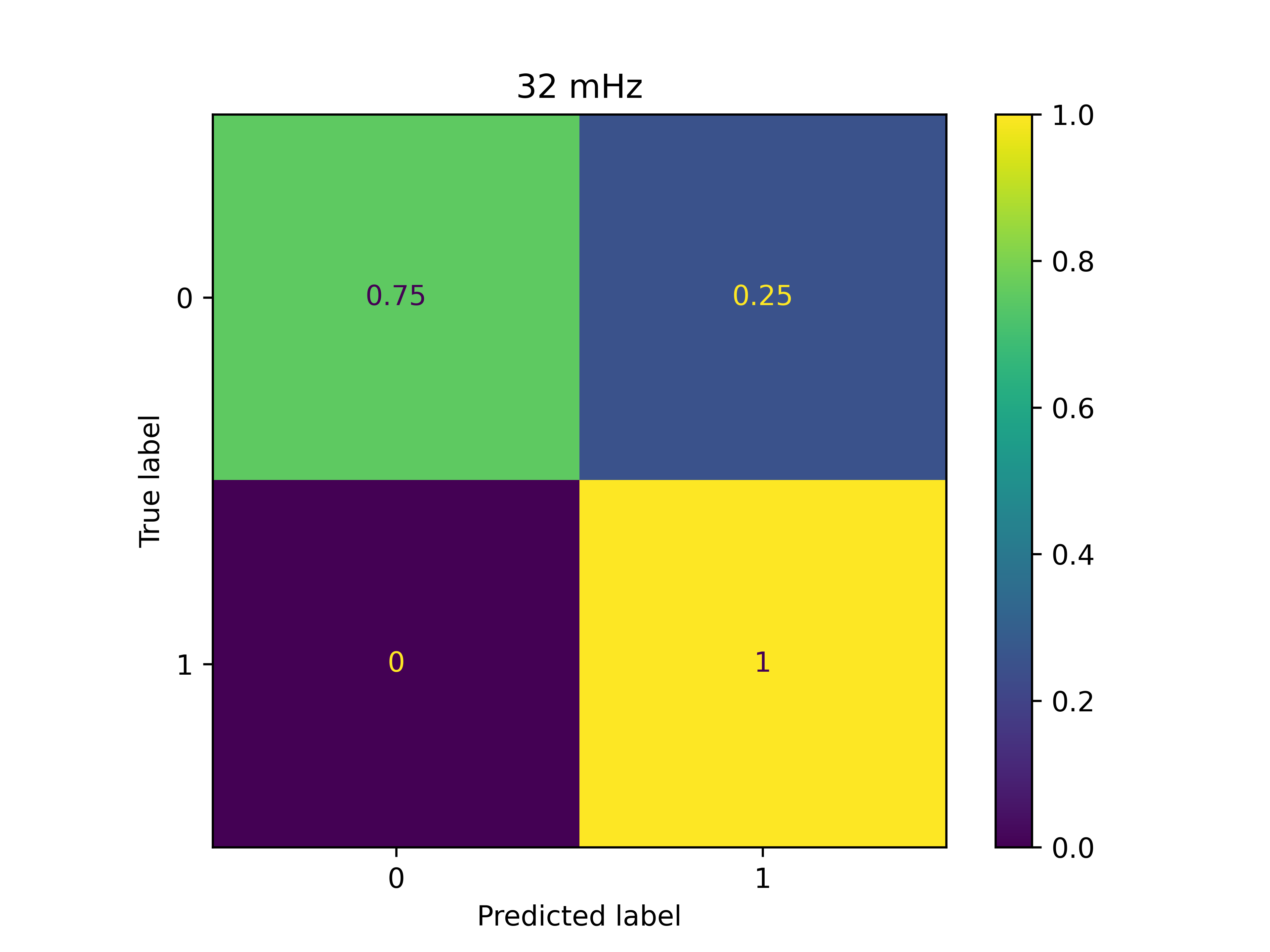}
  \caption{}
  \label{CMVc}
\end{subfigure}

\medskip
\begin{subfigure}{0.33\textwidth}
  \includegraphics[width=\linewidth]{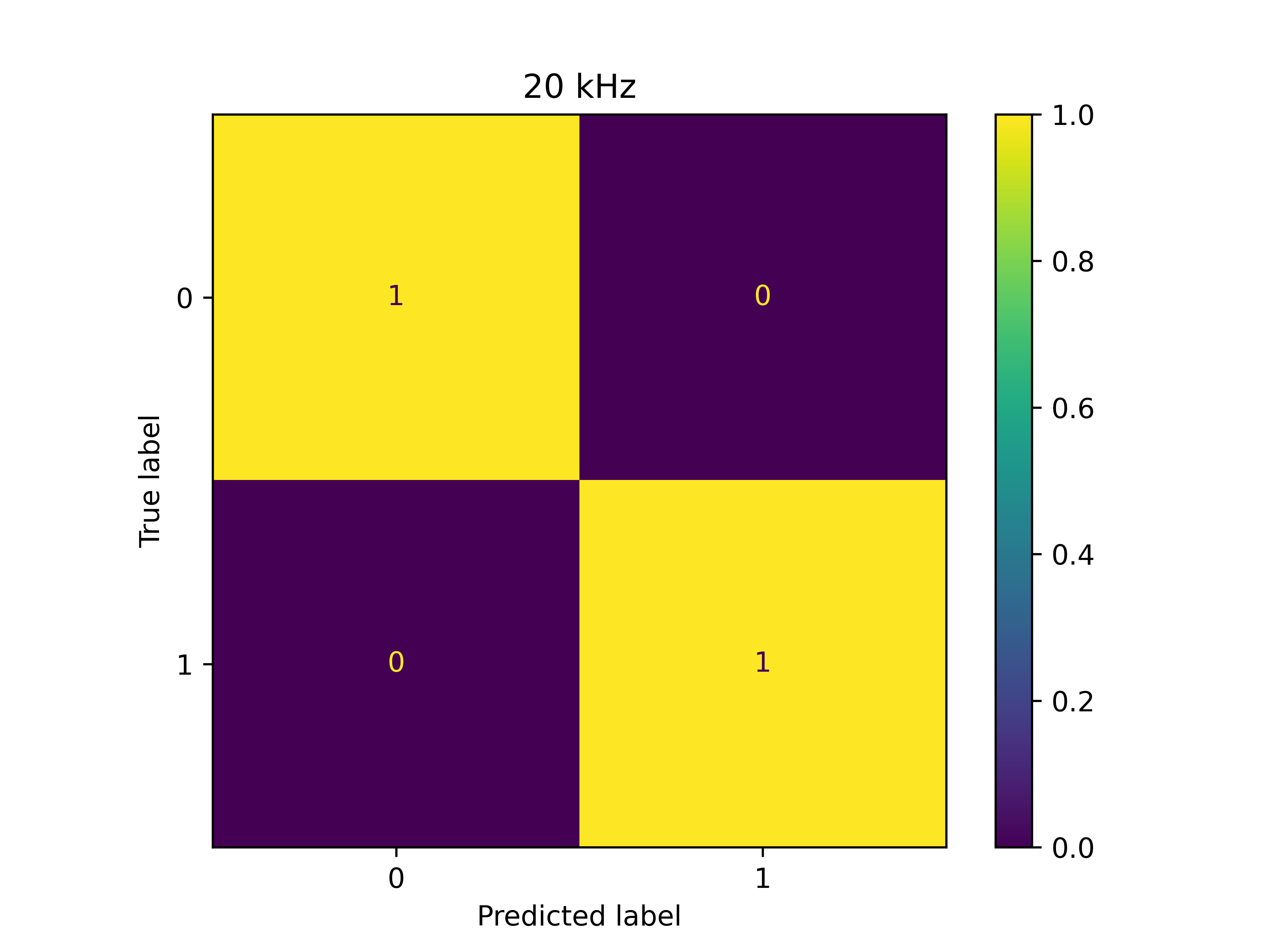}
  \caption{}
  \label{CMIXd}
\end{subfigure}\hfil
\begin{subfigure}{0.33\textwidth}
  \includegraphics[width=\linewidth]{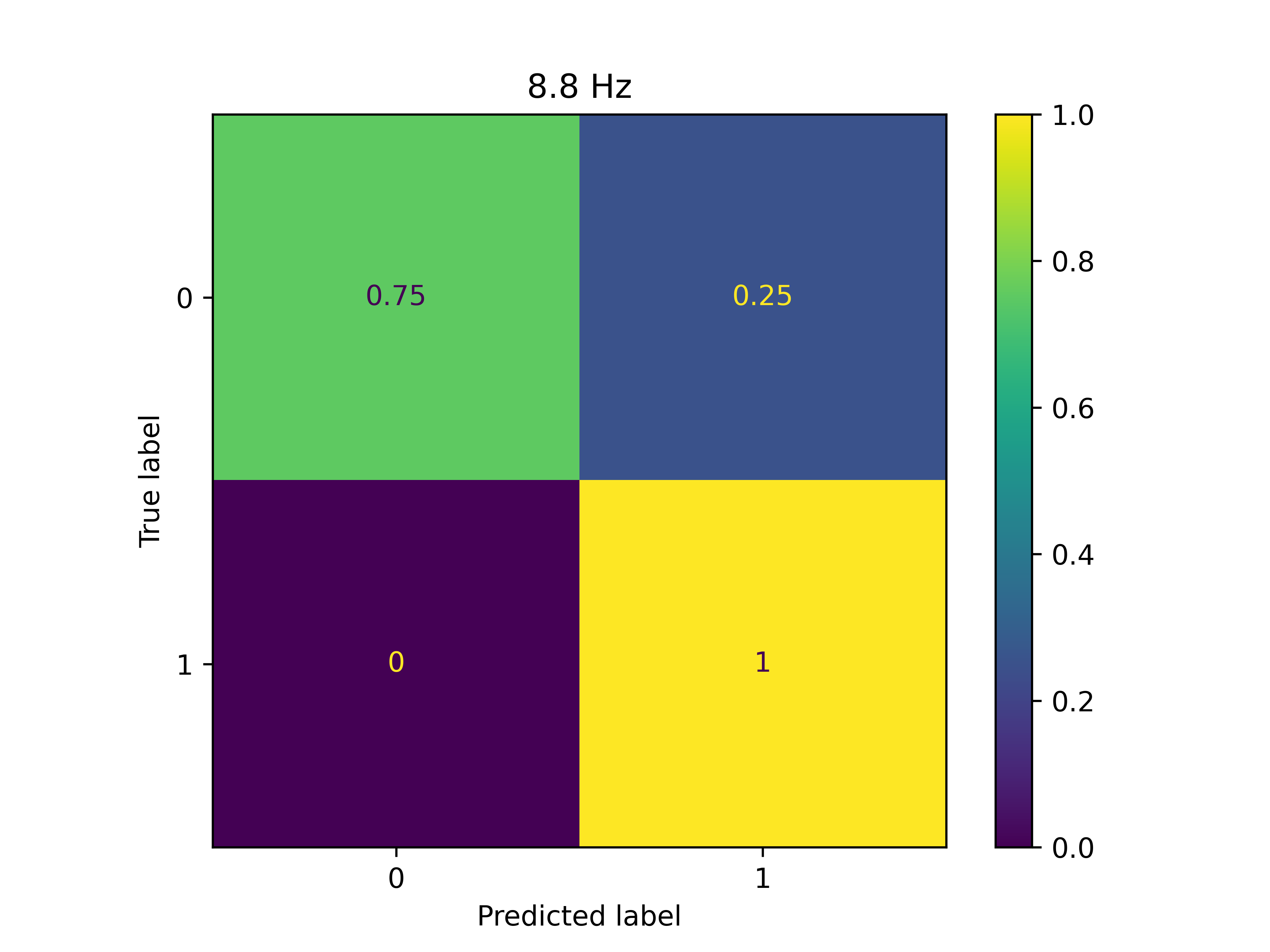}
  \caption{}
  \label{CMIXe}
\end{subfigure}\hfil 
\begin{subfigure}{0.33\textwidth}
  \includegraphics[width=\linewidth]{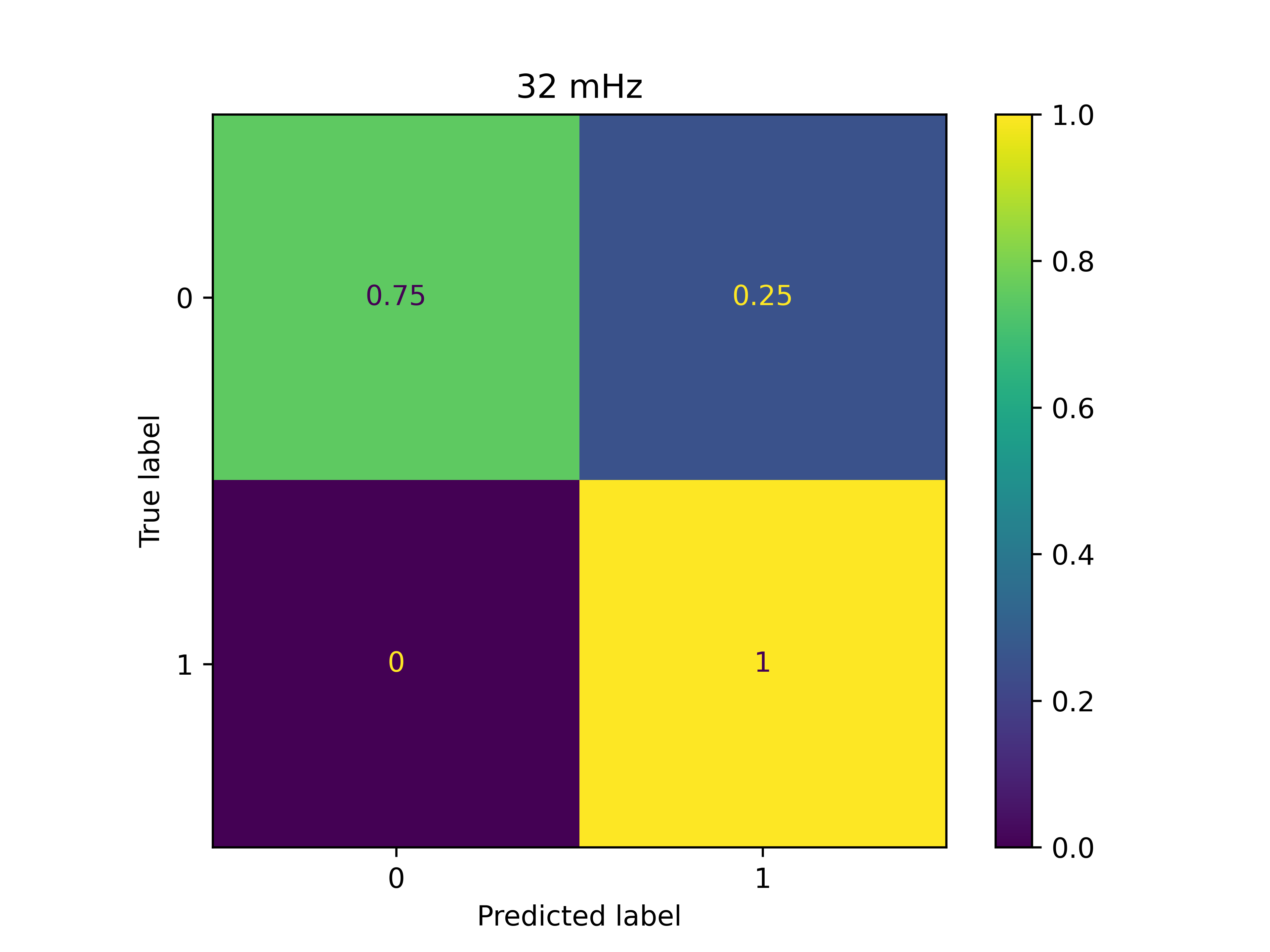}
  \caption{}
  \label{CMIXf}
\end{subfigure}
\caption{Classification confusion matrix for EIS data after fully charging and resting for 15 minutes (a-c) and after fully discharging and resting for 15 minutes (d-f) with a linear support vector machine algorithm. Label 1 corresponds to batteries which show a discharge capacity greater than 80$\%$ of its initial capacity at cycle 200, while label 0 corresponds to a discharge capacity less than 80$\%$ at cycle 200. }
\label{CM}
\end{figure}

Our linear SVM model is able to accurately classify all high-performing and low-performing batteries when trained on only the 20 kHz impedance response from the first 20 cycles for both states of charge (SOC) (V) shown in Figure \ref{CMVa} and (IX) shown in Figure \ref{CMIXd}. SOC (V) corresponds to batteries which are fully charged and rested for 15 minutes. SOC (IX) corresponds to batteries which are fully discharged and rested for 15 minutes.  Furthermore, this accuracy at 20kHz holds even at electrochemically non-stable states such as (III) where measurements were taken after 20 minutes of charging. At all three states of charge, given only the 20 kHz impedance response, all 60 data points are correctly classified. 

While 20 kHz gives ideal predictions, other lower frequencies such as 8.8 Hz and 32 mHz, also give good predictions at all three states of charge, shown in Figure  \ref{CMVb},  \ref{CMVc}, \ref{CMIXe}, and \ref{CMIXf}. Here, all data points corresponding to high-performing batteries are correctly classified, but some data points corresponding to low-performing batteries are incorrectly classified as high-performing.

\begin{figure}[htb]
    \centering 
\begin{subfigure}{0.33\textwidth}
  \includegraphics[width=\linewidth]{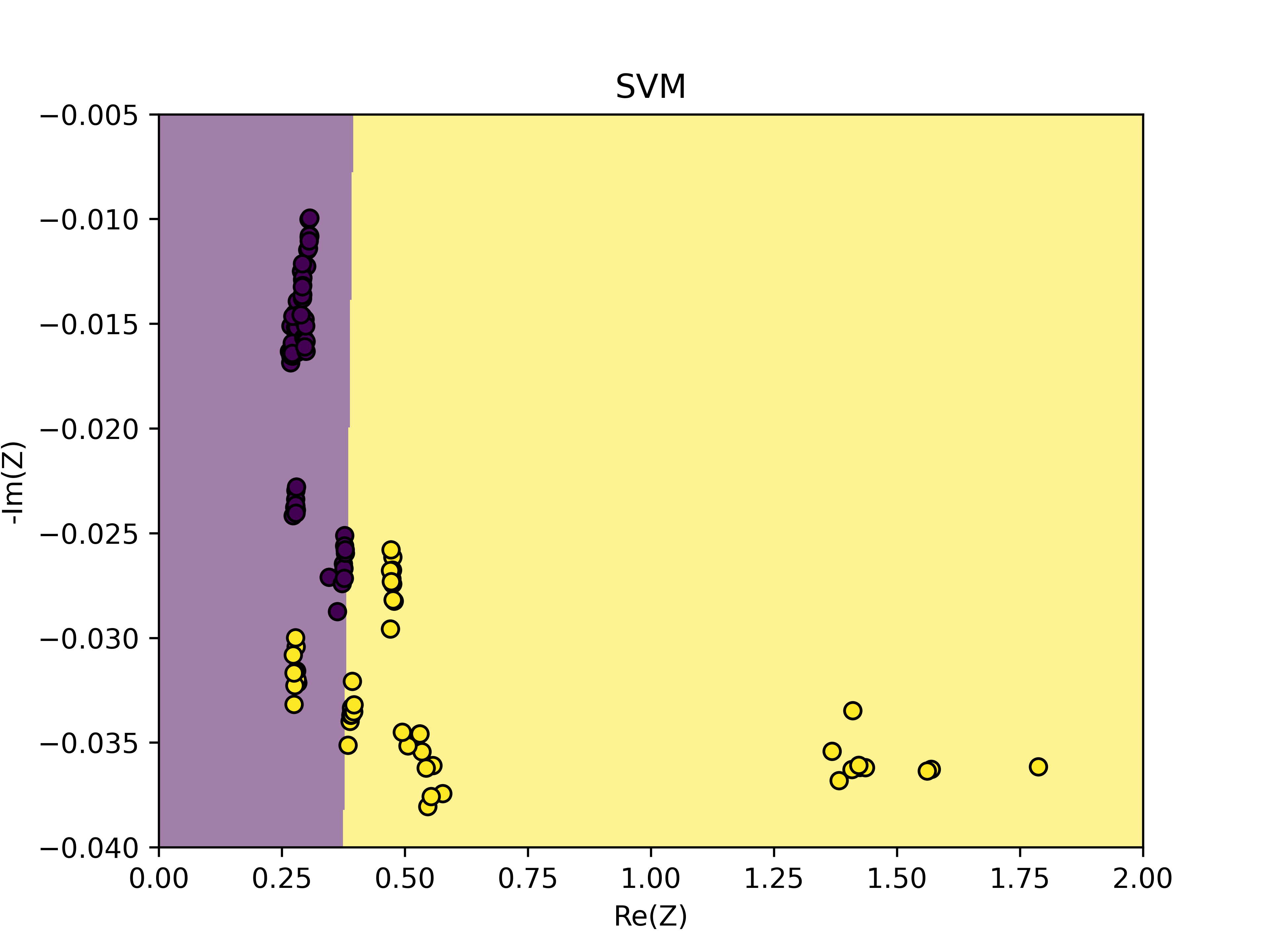}
  \caption{}
  \label{Moda}
\end{subfigure}\hfil
\begin{subfigure}{0.33\textwidth}
  \includegraphics[width=\linewidth]{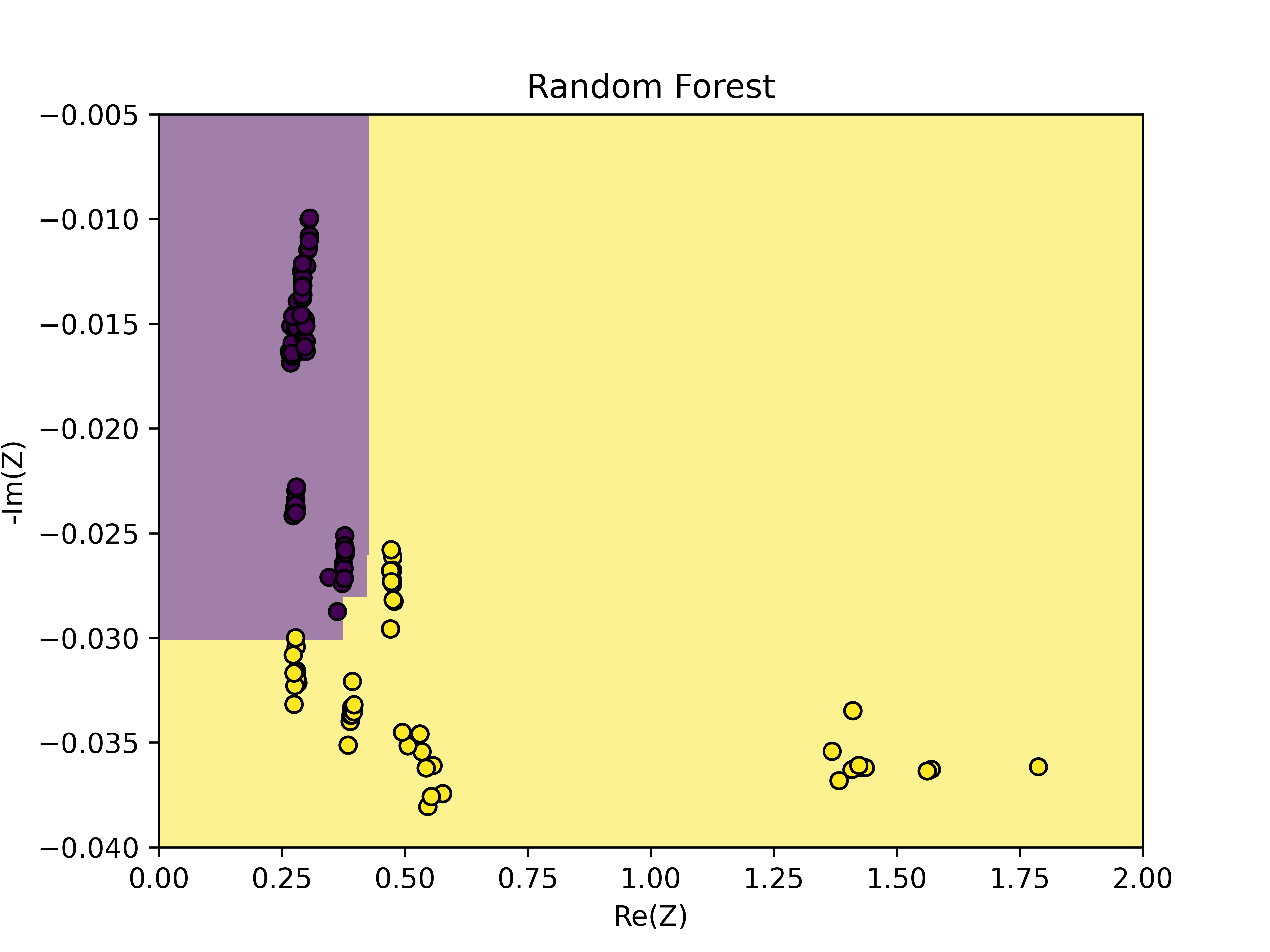}
  \caption{}
  \label{Modb}
\end{subfigure}\hfil
\begin{subfigure}{0.33\textwidth}
  \includegraphics[width=\linewidth]{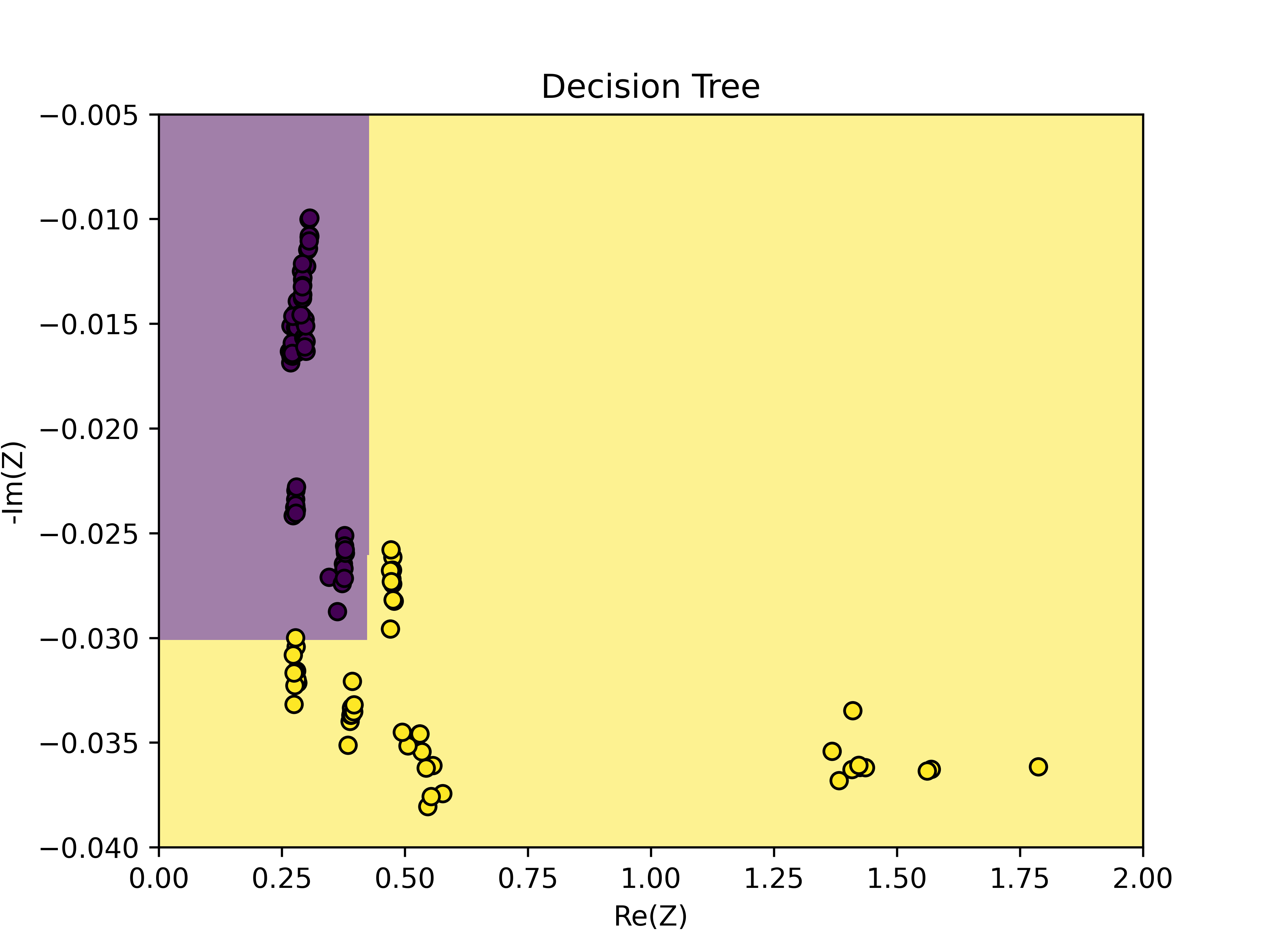}
  \caption{}
  \label{Modc}
\end{subfigure}

\medskip
\begin{subfigure}{0.33\textwidth}
  \includegraphics[width=\linewidth]{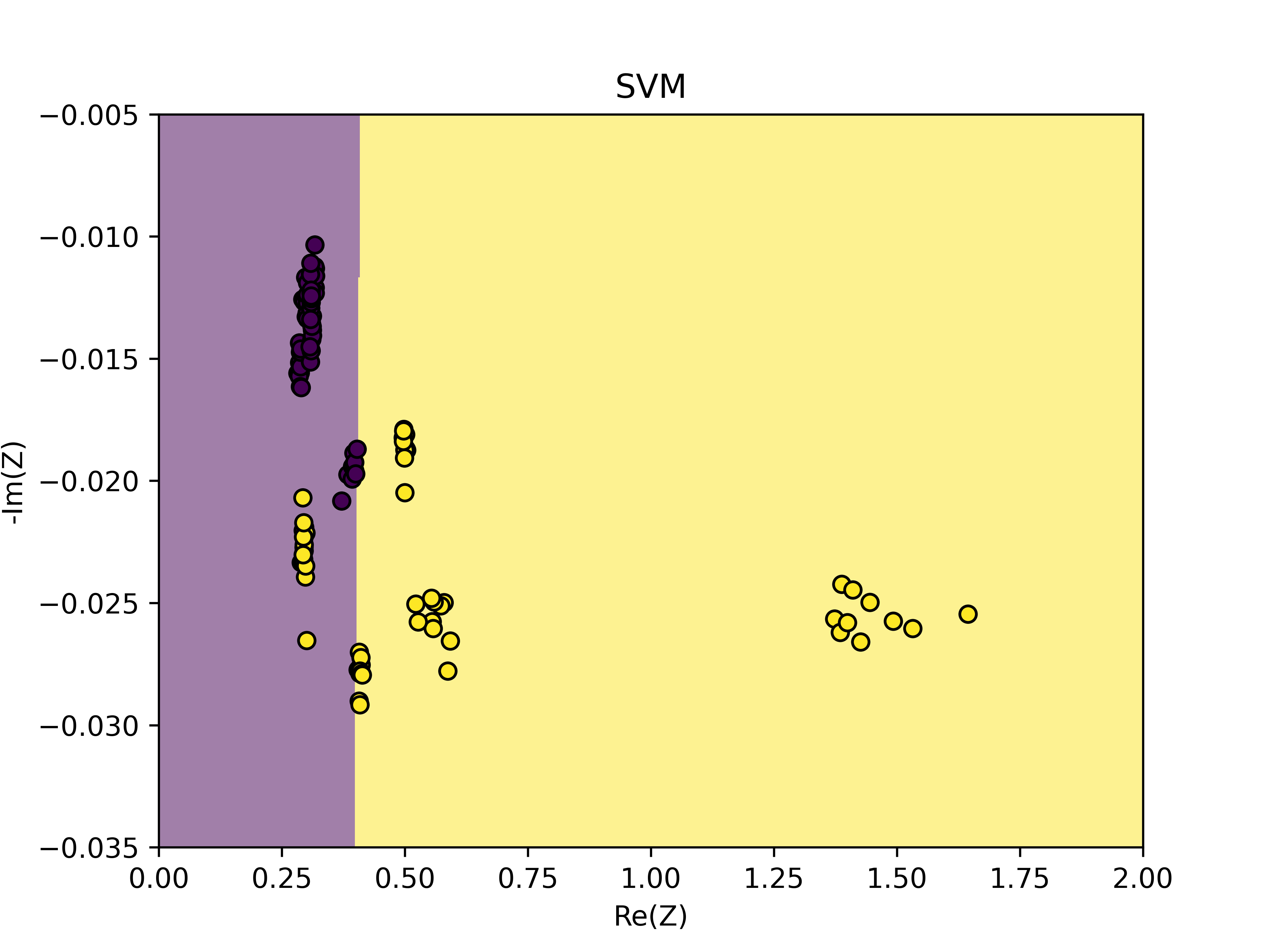}
  \caption{}
  \label{Modd}
\end{subfigure}\hfil
\begin{subfigure}{0.33\textwidth}
  \includegraphics[width=\linewidth]{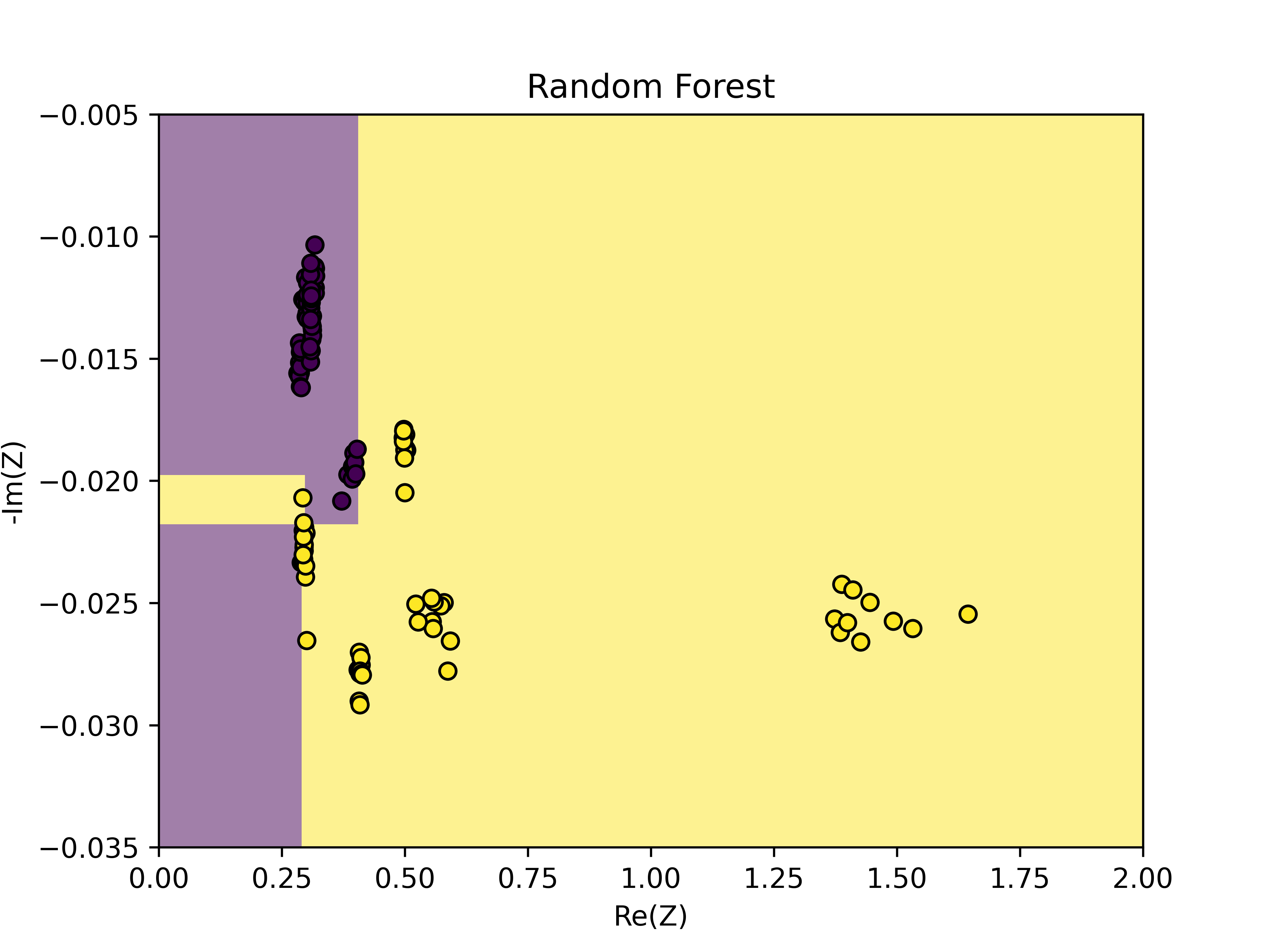}
  \caption{}
  \label{Mode}
\end{subfigure}\hfil 
\begin{subfigure}{0.33\textwidth}
  \includegraphics[width=\linewidth]{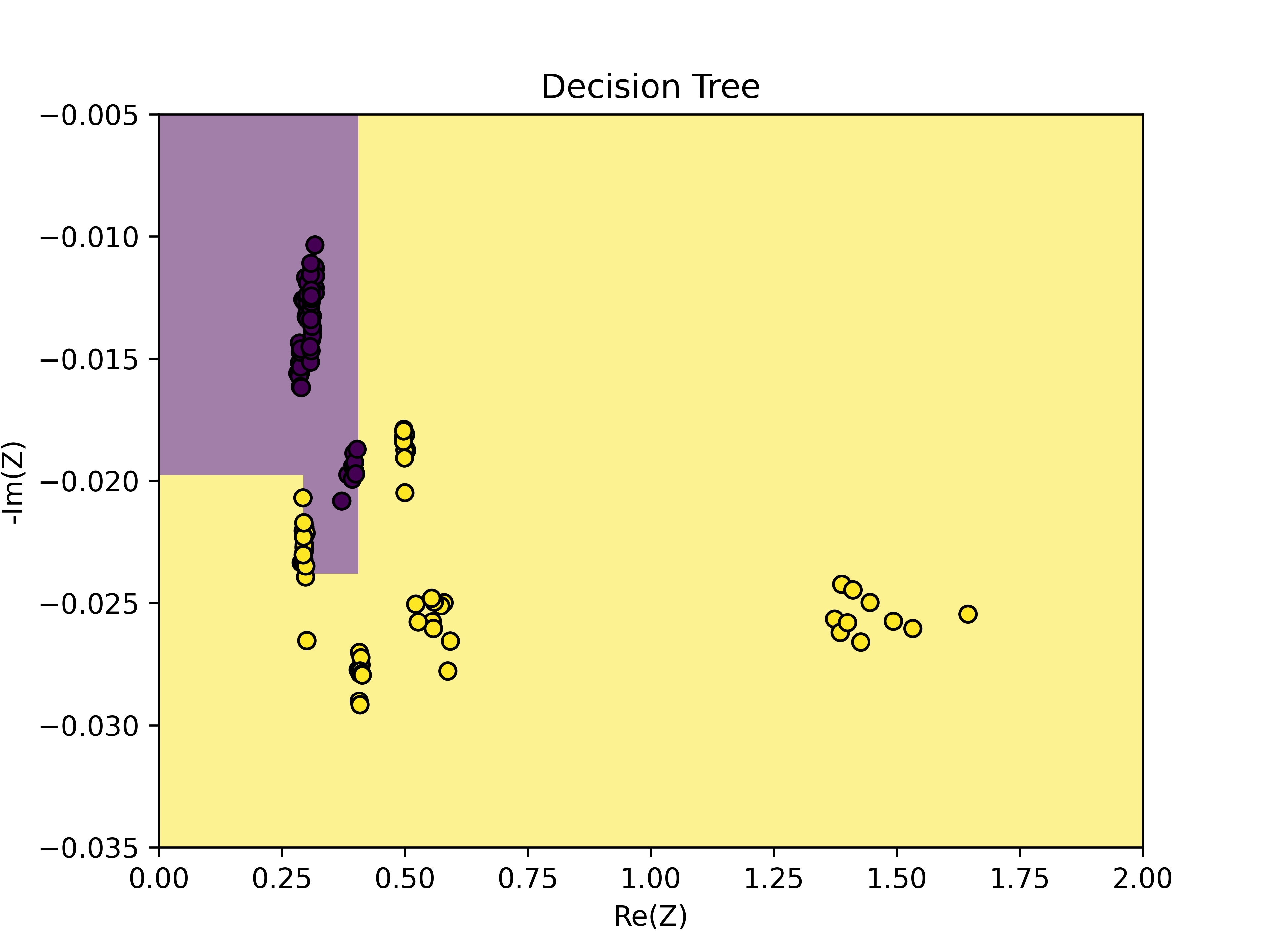}
  \caption{}
  \label{Modf}
\end{subfigure}
\caption{Classification decision boundary for EIS data after fully charging and resting for 15 minutes (a-c) and after fully discharging and resting for 15 minutes (d-f). a and d corresponds to our proposed Support Vector Machine model, b and e corresponds to a random forest classifier, and c and f corresponds to a decision tree classifier. Yellow corresponds to batteries which show a discharge capacity greater than 80$\%$ of its initial capacity at cycle 200, while purple corresponds to a discharge capacity less than 80$\%$ at cycle 200. }
\label{Mods}
\end{figure}

Due to the linear separation of classes and relatively small number of independent training data points, we chose a linear SVM maximum-margin algorithm for its robustness and good generalisation ability \cite{awad2015support}. As the number of independent training data points increase with additional published data it may become more beneficial to instead use another algorithm such as random forest or decision tree classifiers which are prone to over-fitting when the number of data points is not very large. Figure \ref{Mods} shows the decision boundary for SVM, random forest, and decision tree algorithms when trained on the 20 kHz Impedance response for all batteries in the Zhang dataset. Here, we can see that data points corresponding to low-performing (purple) and high-performing (yellow) cells are well separated by quadrants.

\subsection{Remaining Useful Life Prediction}

Next, we propose and test a method to predict the remaining useful life (RUL) of both high and low-performing batteries at early cycle numbers. For RUL predictions, we drop 25C04 from our training set and move 25C07 from the testing to training set, because the cycle of failure for 25C04 cannot be determined without extrapolation. 

\begin{figure}[htb]
    \centering 
\begin{subfigure}{0.48\textwidth}
  \includegraphics[width=\linewidth]{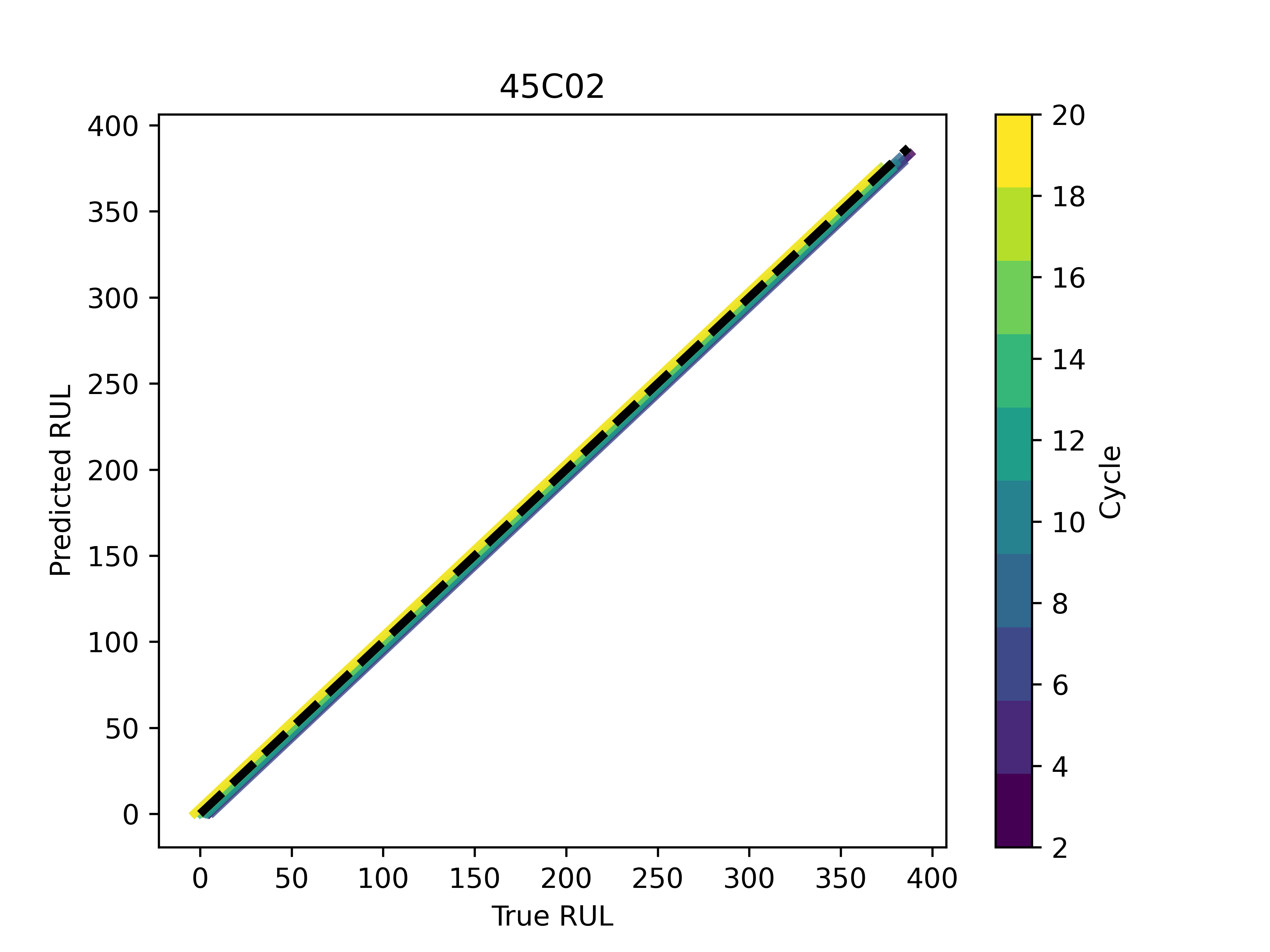}
  \caption{}
  \label{VRa}
\end{subfigure}
\medskip

\begin{subfigure}{0.48\textwidth}
  \includegraphics[width=\linewidth]{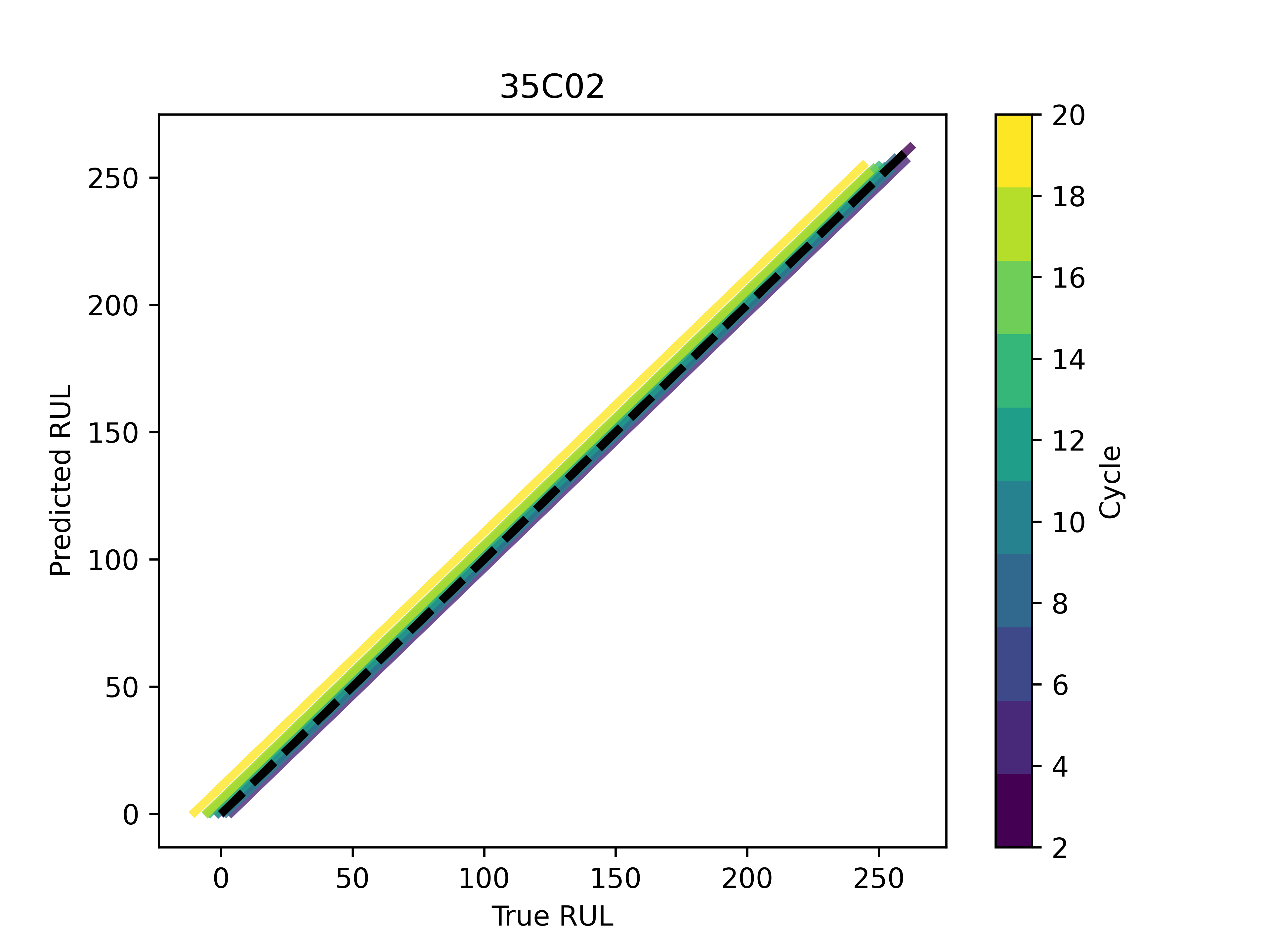}
  \caption{}
  \label{VRb}
\end{subfigure}\hfil
\begin{subfigure}{0.48\textwidth}
  \includegraphics[width=\linewidth]{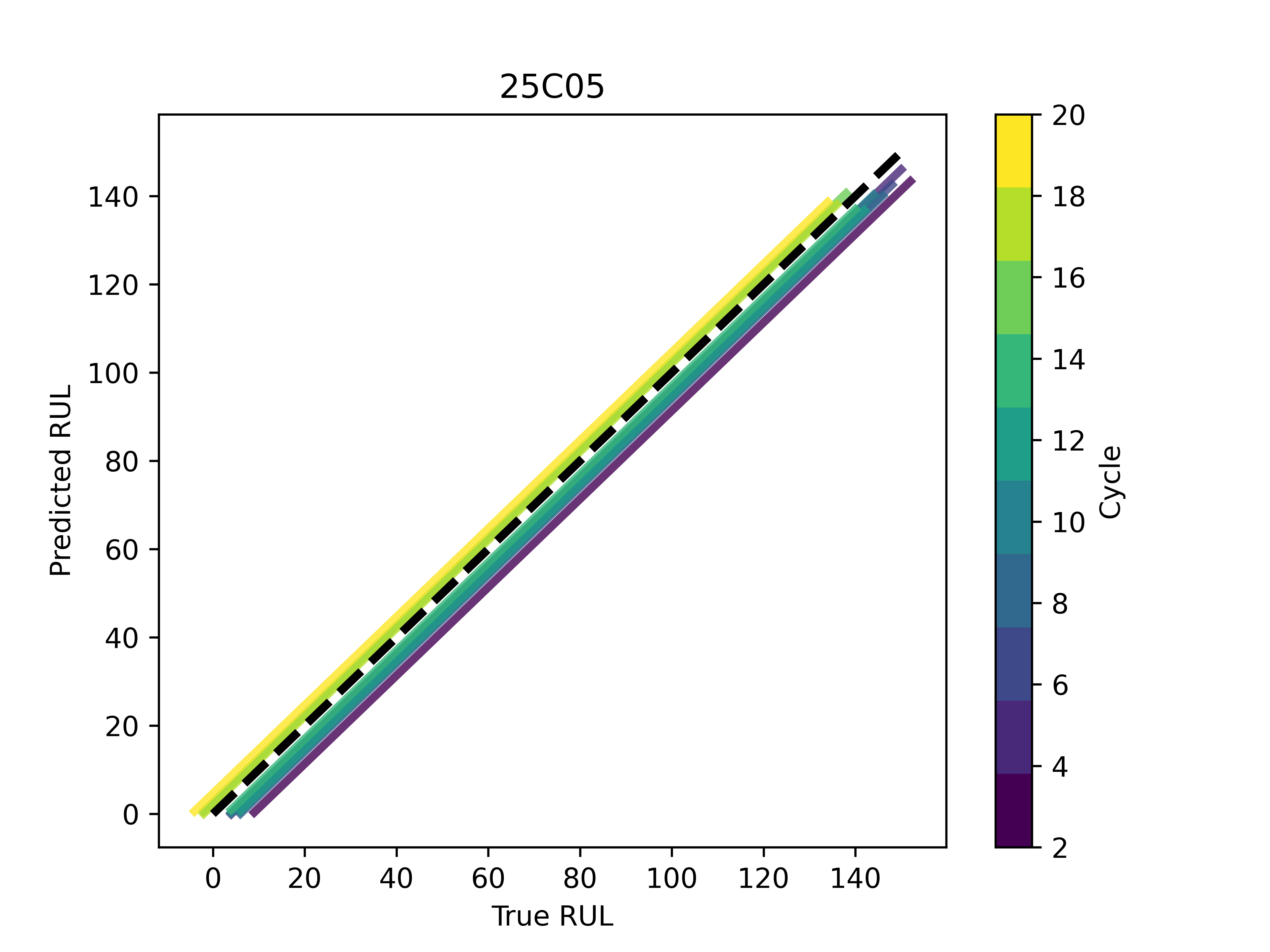}
  \caption{}
  \label{VRc}
\end{subfigure}
\medskip

\begin{subfigure}{0.48\textwidth}
  \includegraphics[width=\linewidth]{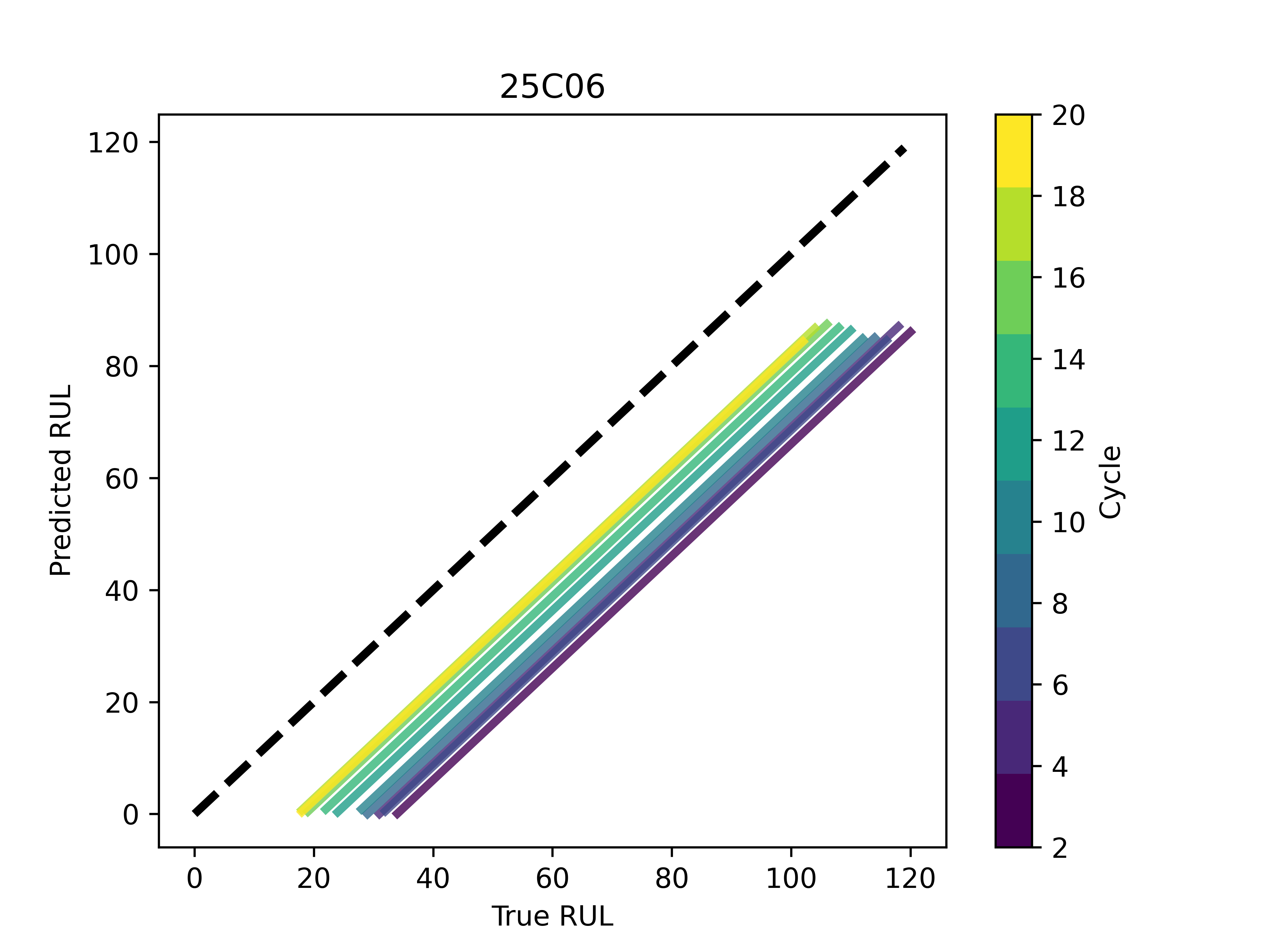}
  \caption{}
  \label{VRd}
\end{subfigure}\hfil 
\begin{subfigure}{0.48\textwidth}
  \includegraphics[width=\linewidth]{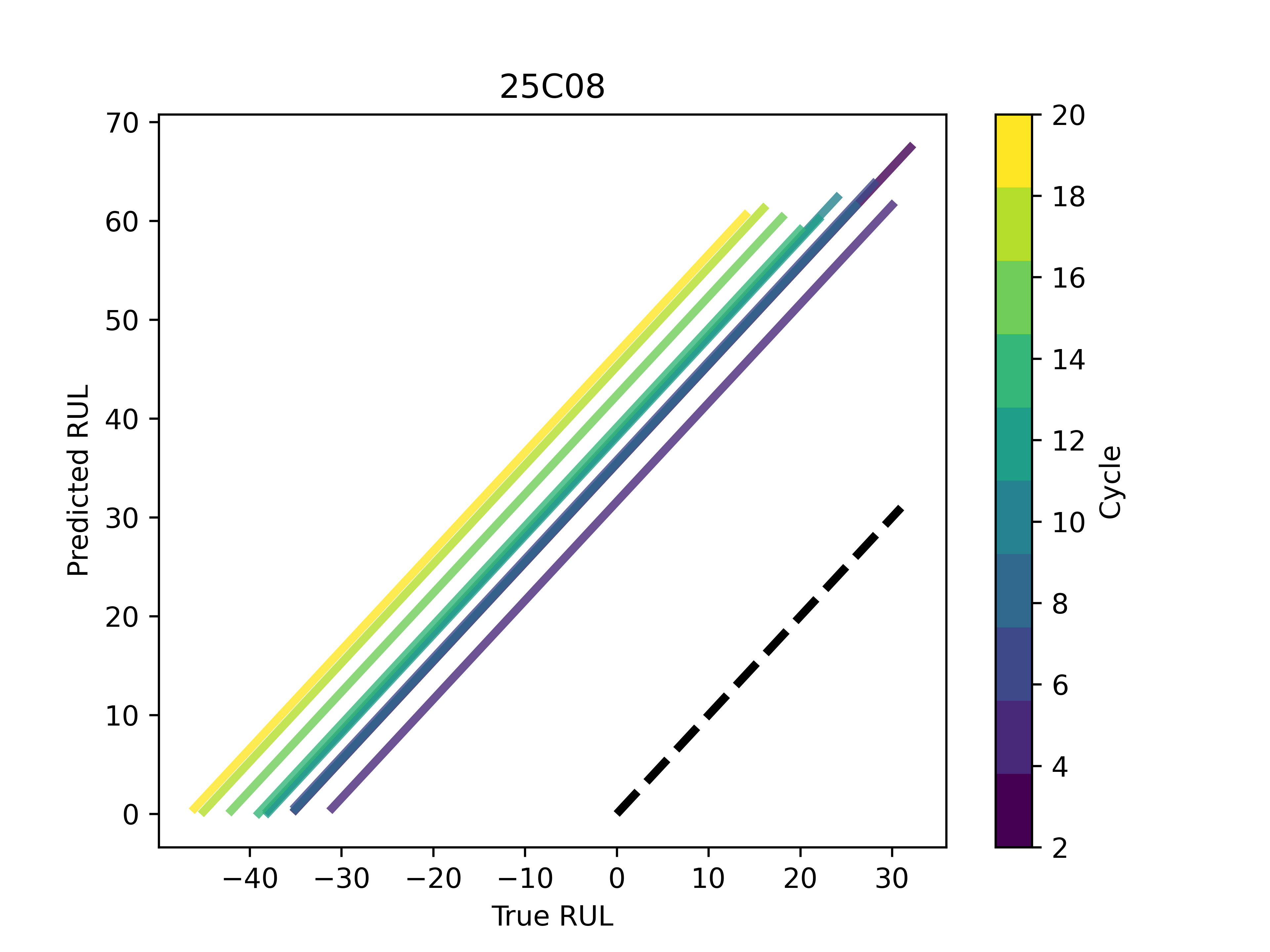}
  \caption{}
  \label{VRe}
\end{subfigure}
\caption{RUL Lasso regression prediction after fully charging and resting for 15 minutes at temperatures (a) (a) 45 \textdegree C, (b) 35 \textdegree C, and (c-e) 25 \textdegree C. The black dashed line represents perfect agreement between prediction and reality. The lasso regression model is trained on -Im(Z) at 20 kHz and 8.8 Hz and temperature. The model predicts the RUL at the current cycle, within the first 20 cycles, and we extrapolate with cycling to failure to create the lines seen here.}
\label{VRegr}
\end{figure}

We train a lasso regression model with a linear kernel on the negative out-of-phase 20 kHz and 8.8 Hz impedance response (-Im(Z)) as well as temperature for the first 20 cycles of batteries in the training set.  Figure \ref{VRegr} shows the RUL predictions for the first 20 cycles of batteries in the testing set, extrapolated to predicted failure. Our model is able to accurately predict the RUL of high performing batteries 45C02, 35C02, and the best low-performing battery 25C05. The predictions lose accuracy for the lowest performing batteries as seen in Fig \ref{VRegr} (d,e). However, we can still determine that 25C06 and 25C08 will be the second worst and worst performing batteries respectively.

\begin{figure}[htb]
    \centering 
\begin{subfigure}{0.48\textwidth}
  \includegraphics[width=\linewidth]{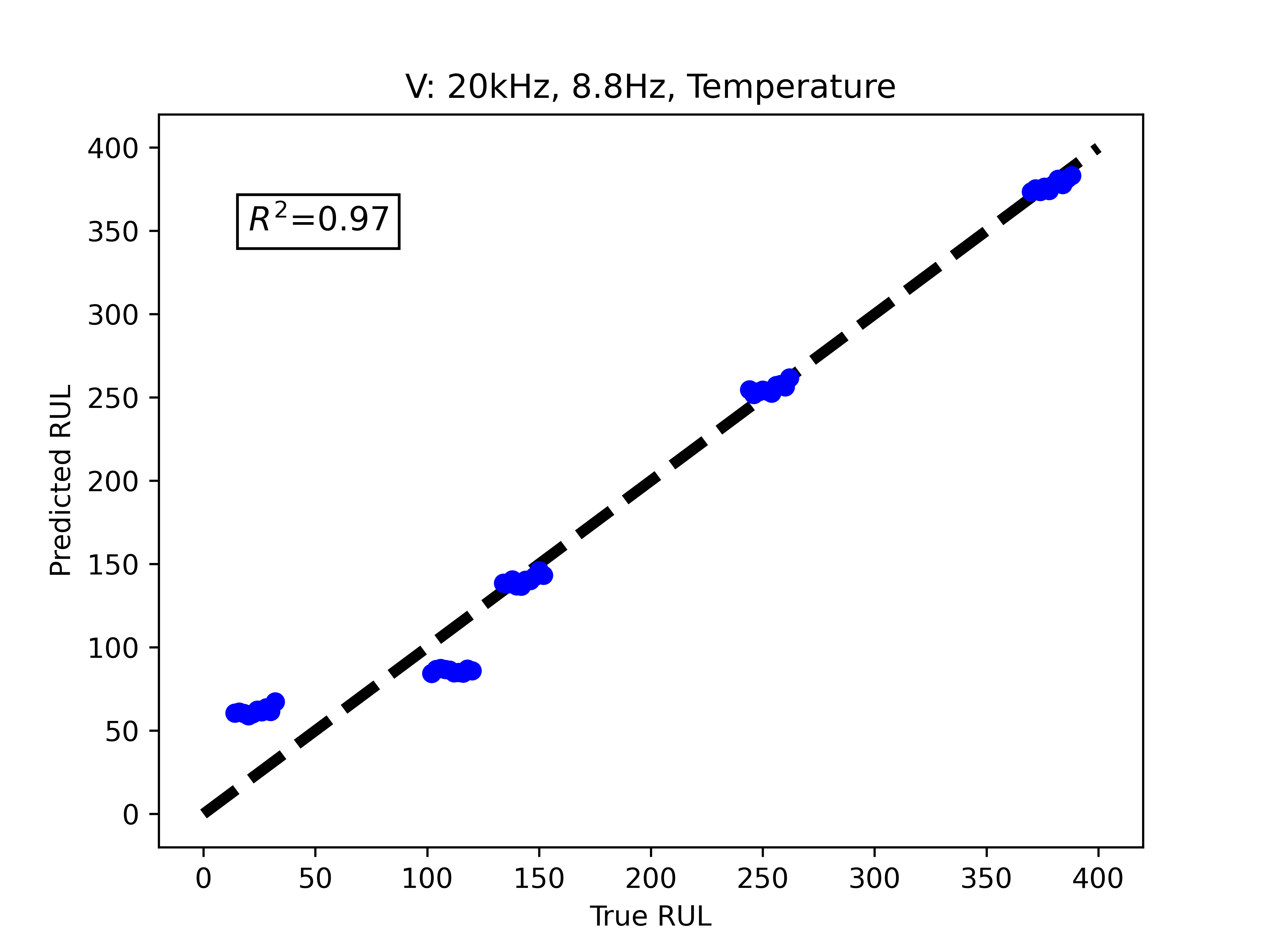}
  \caption{}
  \label{Ra}
\end{subfigure}\hfil
\begin{subfigure}{0.48\textwidth}
  \includegraphics[width=\linewidth]{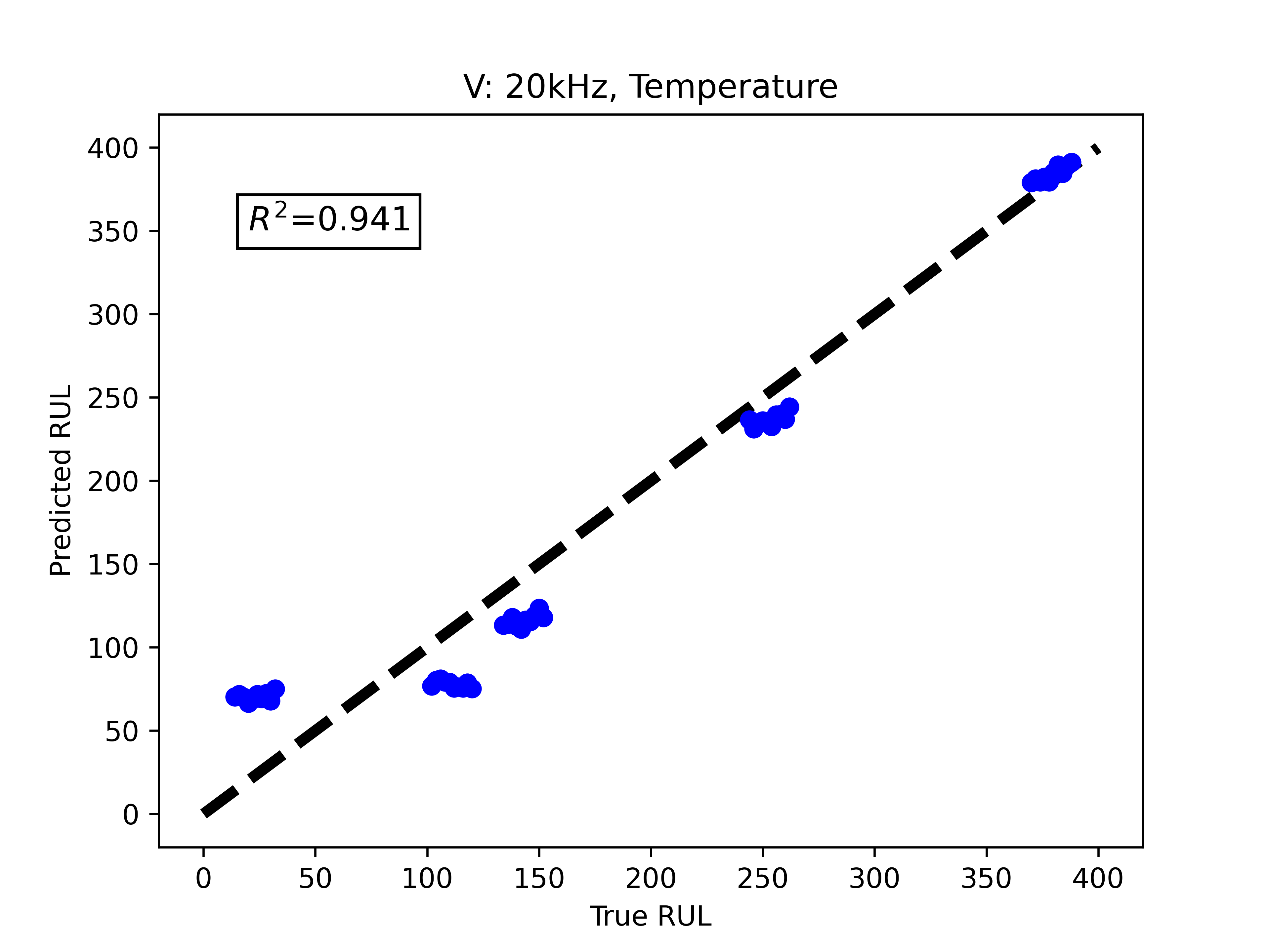}
  \caption{}
  \label{Rb}
\end{subfigure}
\medskip

\begin{subfigure}{0.48\textwidth}
  \includegraphics[width=\linewidth]{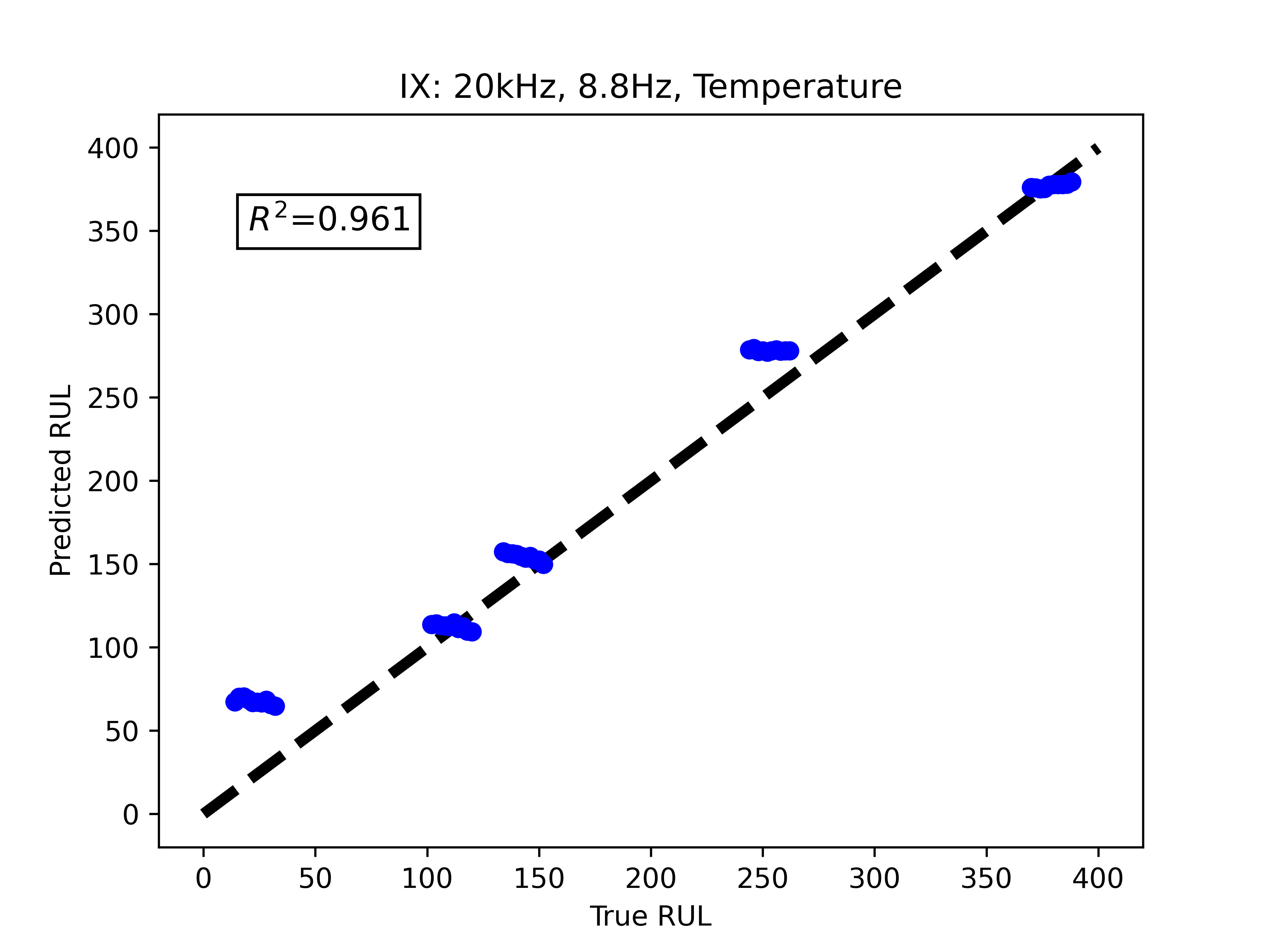}
  \caption{}
  \label{Rc}
\end{subfigure}\hfil
\begin{subfigure}{0.48\textwidth}
  \includegraphics[width=\linewidth]{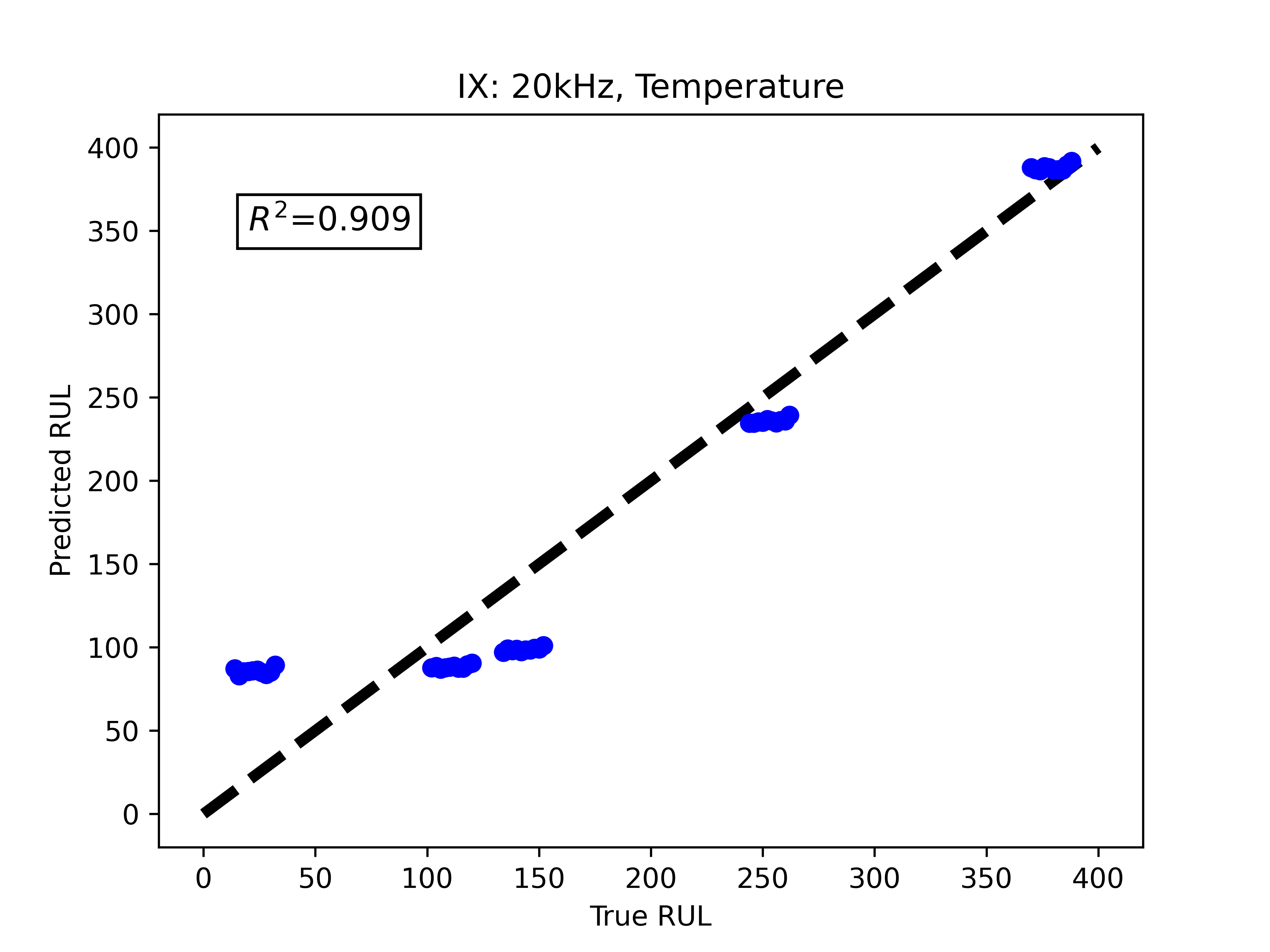}
  \caption{}
  \label{Rd}
\end{subfigure}
\medskip

\begin{subfigure}{0.48\textwidth}
  \includegraphics[width=\linewidth]{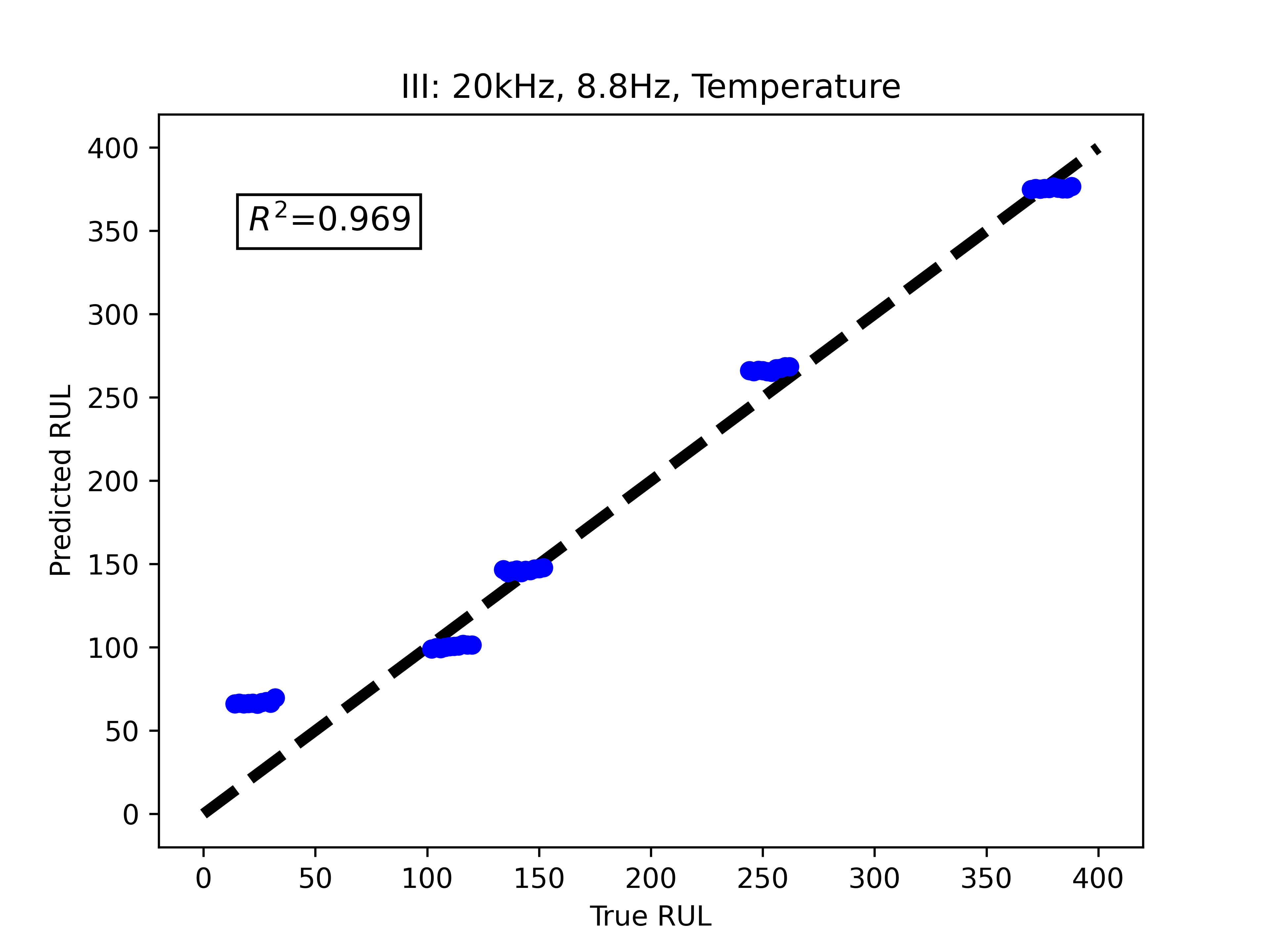}
  \caption{}
  \label{Re}
\end{subfigure}\hfil
\begin{subfigure}{0.48\textwidth}
  \includegraphics[width=\linewidth]{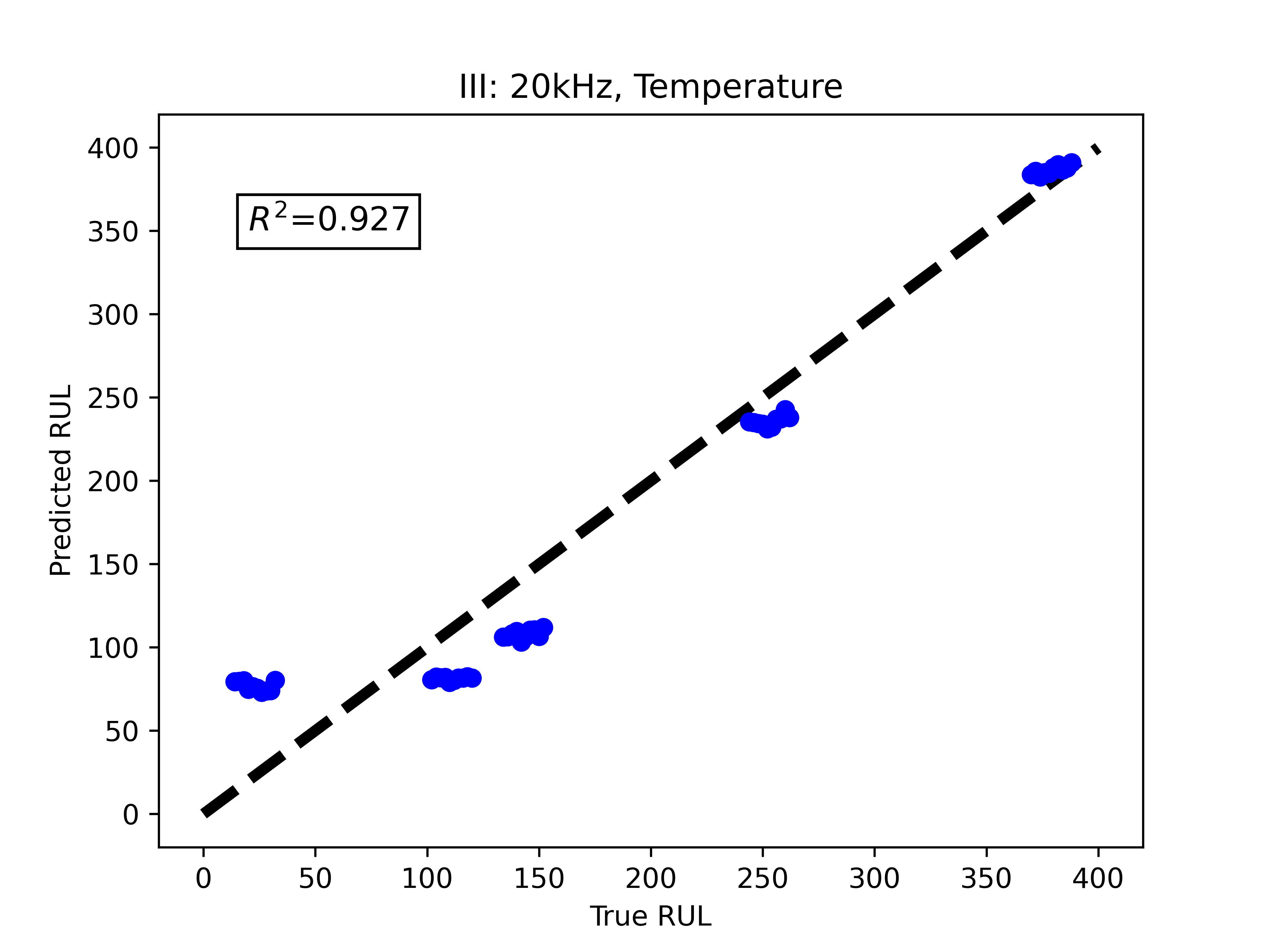}
  \caption{}
  \label{Rf}
\end{subfigure}
\caption{RUL Lasso regression predictions trained on -Im(Z) at 20 kHz and 8.8 Hz and temperature (a,c,e) and trained on -Im(Z) at 20 kHz and temperature (b,d,f). (a-b) correspond to an SOC of fully charged and resting for 15 minutes (V), (c-d) is fully discharged and resting for 15 minutes (IX), and (e-f) is 20 minutes of charging (III). The black dashed line represents perfect agreement between prediction and reality. The $R^2$ coefficient of determination for each model is shown as an inset on the top left.}
\label{R2Reg}
\end{figure}

Figure \ref{R2Reg} shows the stability of our model at stable (V and IX) and non-stable (III) states of charge. For all three states of charge, we can accurately predict the RUL at early cycles when trained on negative out-of-phase impedance response (-Im(Z)) at 20 kHz and 8.8 Hz and temperature of the first 20 cycles. Figure \ref{R2Reg} (a-c) show that the $R^2$ coefficient of determination is larger than 0.96 for all three states of charge. Figure \ref{R2Reg} (d-f) shows RUL predictions when trained on only the negative out-of-phase impedance response (-Im(Z)) at 20 kHz and temperature of the first 20 cycles. With this simplified model, we achieve $R^2$ greater than 0.9 for all three states of charge. 

Additionally, if the model is trained on only the negative out-of-phase impedance response (-Im(Z)) at 8.8 kHz and temperature or on the negative out-of-phase impedance response (-Im(Z)) at 32 mHz and temperature, we are able to achieve an $R^2$ greater than 0.89 for all three states of charge. Notably, our choices of frequency perform better than 17.80 Hz and 2.16 Hz which were the notable features from ARD in Zhang's original work on this dataset.

\clearpage

\section{Methods}
\subsection{Data Processing}
All data used in this analysis comes from the public repository associated with Ref. \cite{Zhang2020degradation}. We find the discharge capacity for a given cycle of a battery by taking the maximum value of capacity for 'ox/red'$=0$, which corresponds to the discharge process. RUL is determined as the number of cycles until a discharge capacity of less than $0.8*C_1$ is measured, where $C_1$ is the discharge capacity at cycle 1. For battery 25C04, the RUL cannot be directly measured, as the battery was not cycled until the failure condition. Here, we use linear extrapolation to find an estimated RUL of 114 at cycle 2, which is well below the 200 cycle boundary between classes 0 and 1.

In the original dataset, capacity is measured every odd cycle, while EIS is measured every even cycle. In this work, we associate the discharge capacity from the previous cycle with the EIS measurement of the current cycle, i.e. the capacity measured during the first cycle is associated with the EIS at cycle 2.  

\subsection{Support Vector Machine}
Support vector machine (SVM)  \cite{smola2004tutorial,awad2015support,bishop2006pattern,chang2011libsvm} classification is used for problems which may not be completely separable by a hyperplane. We allow some samples in the training set to be incorrectly classified with a penalty in the minimization function, known as maximum margin classification. The linear ($\phi =$ Identity function) SVM classifier used in this work finds hyperplane parameters $w$ and $b$ that minimize the following loss function:

\begin{equation}
\frac{1}{2}w^Tw + C \sum_{i=0}^n max(0,1-y_i(w^T\phi(x_i)+b))
\end{equation}

where $C$ is the strength of the penalty, $x_i$ is training vector $i$, and $y_i=\pm 1$ is the true class for sample $i$. Intuitively, the second term is zero for correct predictions outside of the margin boundary, and non-zero for incorrect predictions or correct predictions inside the margin boundary. 

\subsection{Decision Tree and Random Forest}
For both models, we use the Gini impurity \cite{gini1912variabilita, menze2009comparison} to measure the quality of the decision split. The Gini impurity for data at a split m ($Q_m$) is given by:
\begin{equation}
H(Q_m)= \sum_k p_{mk}(1-p_{mk})
\end{equation}

where $p_{km}$ is the proportion of class k observations in mode m. 

The decision tree classifier \cite{dumont2009fast,song2015decision} finds a split $\theta$ which minimizes the following function:
\begin{equation}
\frac{n_m^{left}}{n_m}H(Q_m^{left}(\theta)) +  \frac{n_m^{right}}{n_m}H(Q_m^{right}(\theta)) 
\end{equation}

where $n_m$ represents the number of samples at node m and left and right refer to the sides of the partition defined by $\theta$.

Random forest classifier \cite{breiman2001random} uses the same technique, but averages over multiple decision trees from sub-samples of the full training set, to improve potential over-fitting. 

\subsection{Lasso}
The objective of the Lasso \cite{friedman2010regularization,kim2007interior} algorithm is minimization of the least squares difference and the $l_1$ prior regularizer of the coefficient vector. 
\begin{equation}
\frac{1}{2n}*||y-Xw||^2_2 + \alpha ||w||_1
\end{equation}
Where the $l_1$ and $l_2$ prior regularizers are defined as:
\begin{equation}
||\Psi||_1 = \sum_{i=0}^n |\Psi_i|
\end{equation}
\begin{equation}
||\Psi||_2^2 = \sum_{i=0}^n \Psi_i^2
\end{equation}
where w is a vector of the model coefficients, n is the number of samples, y is the true value, Xw is the predicted value from the model. 
For SOC V, we find the weights $w = [-2139.97, -318.65, 9.51]$ for parameters $X_i = $\{-Im[Z$_i$(20 kHz)], -Im[Z$_i$(8.8 Hz)], T$_i$\}, respectively.

\section{Discussion}

In this paper, we demonstrate that we can accurately classify lithium-ion batteries as high or low-performing and accurately predict the RUL of the batteries at early cycles with knowledge of only the high frequency (20 kHz) electrochemical impedance response and the temperature of the cell. Classification with mid (8.8 Hz) and low (32 mHz) frequency also shows promising results. We demonstrate that our model can accurately predict the RUL of batteries at early cycles with knowlege of only the 20 kHz impedance response of the system as well as the operating temperature. These RUL predictions can be improved by the addition of a mid frequency (8 Hz) impedance response. The consistent importance of temperature in our RUL models suggests that cycling temperature is closely linked to the RUL of these batteries, and that thermal engineering has the potential to drastically change the lifespan of these batteries \cite{zhang2017remaining, karlsen2019temperature}. Further work at additional temperatures can help find the optimal temperature for battery RUL.  

Previous work \cite{Zhang2020degradation} found that 17.80 Hz and 2.16 Hz were the most relevant frequencies in predicting the RUL of batteries in this dataset, therefore identifying interfacial properties as the primary source of degradation in these batteries. In contrast, our work suggests the viability of frequencies from 20 kHz to 32 mHz. We believe these results prompt the consideration of a more holistic approach to understanding battery degradation, rather than a focus on mid-frequency features. It is important to note that the difference in relevant features may come from the difference in the statement of the problem: in our work we use early cycle EIS data to predict the RUL at only early cycles. The model in Ref. \cite{Zhang2020degradation} predicts the evolution of RUL with cycling. In the second statement of the problem, the model must find parameters which can make reasonable early cycle predictions and which change uniformly with cycle number. We find that the 17.80 Hz and 2.16 Hz impedance response changes relatively linearly with cycle number. We achieve better early cycle predictions using 20 kHz and 8.8 Hz impedance response, and assume that cycle number can be directly measured by the battery management system, rather than being predicted. 

\printbibliography

\end{document}